# Classification of All Blood Cell Images using ML and DL Models

**Rabia Asghar, Sanjay Kumar, Paul Hynds, Abeera Mahfooz**

***Abstract*—** Human blood is primarily composed of plasma, red blood cells, white blood cells, and platelets. It plays a vital role in transporting nutrients to different organs, where it stores essential health-related data about the human body. Blood cells are utilized to defend the body against diverse infections, including fungi, viruses, and bacteria. Hence, the analysis of blood can help physicians in assessing an individual's physiological condition. Blood cells have been sub classified into eight groups: Neutrophils, eosinophils, basophils, lymphocytes, monocytes, immature granulocytes (promyelocytes, myelocytes, and metamyelocytes), erythroblasts, and platelets or thrombocytes on the basis of their nucleus, shape and cytoplasm. Traditionally, pathologists and hematologists in laboratories have examined these blood cells using a microscope before manually classifying them. The manual approach is slower and more prone to human error. Therefore, it is essential to automate this process. In our paper, transfer learning with CNN pre-trained models—VGG16, VGG19, ResNet-50, ResNet-101, ResNet-152, InceptionV3, MobileNetV2 and DenseNet-20 applied to PBC dataset's normal DIB. The overall accuracy achieved with these models lies between 91.375 and /to 94.72%. Hence, inspired by these pre-trained architectures, a model has been proposed to automatically classify the ten types of blood cells with increased accuracy. A novel CNN-based framework has been presented to improve accuracy. The proposed CNN model has been tested on PBC dataset normal DIB. The outcomes of the experiments demonstrate that our CNN-based framework designed for blood cell classification attains an accuracy of 99.91% on the PBC dataset. Our proposed convolutional neural network model performs competitively when compared to earlier results reported in literature.

## I. INTRODUCTION

Blood is a specific type of circulating connective fluid that takes oxygen from the lungs and transports it to all human body cells. The cells require oxygen for metabolism, which the blood carries from the lungs to the cells. In this way, blood cells carry hormones, and eliminates unnecessary materials that are eventually eliminated by organs like the liver, kidneys, or intestine. The blood is comprised of plasma that forms liquid portion and cell fragments. The cell fragments are composed of white blood cells (WBCs) with 1% proportion, responsible for immunity. Red blood cells (RBCs) comprise 40-50% of the total blood volume, carrying oxygen and carbon dioxide, whereas platelets have the crucial role of promoting blood clotting [1,2]. WBCs can be divided into two general groups based on the presence of granules: granulocytes and agranulocytes (non-granulocytes). Lymphocytes and monocytes fall under agranulocytes, while neutrophils, eosinophils, and basophils are considered to be granulocytes. Undeveloped WBCs called immature granulocytes (IG) are expelled from the bone marrow into the blood. The presence of immature granulocytes (promyelocytes, myelocytes, and metamyelocytes) in the blood signifies an early reaction to an infection, swelling, or some other type of problem with the bone marrow like leukemia, except the blood from newly born children or pregnant women. White blood cells act as defenders against infections, breaking down foreign proteins present in bacteria, viruses, and fungi. Through division, WBCs fight infections and diseases by recognizing, identifying, and attaching themselves to these foreign antigens [3]. Red blood cells, also known as erythroblasts, help tissues produce energy by delivering appropriate oxygen. When energy is produced, waste in the form of carbon dioxide is also formed. RBCs are responsible for providing that carbon dioxide to the lungs so that it is exhaled. Erythroblasts are immature RBCs that are usually present in the blood of newly born child in a duration from 0-4 months. Their presence in human blood after the neonatal period (0-4 months) indicates severe problems like damaged bone marrow, stress, and malignant tumor that may lead to cancer or benign that grows in size but do not infect other body parts. Platelets are also known as thrombocytes and are essential for the immunity system; their primary responsibility is to stop bleeding. If bleeding starts from an injury or blood vessel damage somewhere in the body, the brain sends an alerting signal to the platelets. The platelets flow to the wounded area, cluster together, and form a clot, sealing the blood vessel to stop the bleeding. They also play an important role in tissue repair and remodeling to prevent tumor progression and leakage of vesicant fluids. They comprise a tiny proportion, i.e., less than 1% of blood volume. Typically, the common rates of neutrophils in the blood are 0-6%. Eosinophils comprise 1-3%, basophils 0-1%, lymphocytes 25-33%, and monocytes 3-10% of the leukocytes floating in the blood [4].

The classification of blood cells is a current research area for scientists trying to diagnose diseases that affect blood cells. Blood cell classification using microscopic images of blood would only be done manually by medical professionals with the necessary experience and training. Blood is analyzed in two different ways. The first method is a complete blood count (CBC) test that calculates the total percentage of RBCs, WBCs, and platelets; the second is the peripheral blood smears (PBS) test. These results represent the patient's overall health. Microscopic blood images can accurately reveal the types of RBCs, WBCs, and platelets, enabling early disease diagnosis. Every type of cell present in human blood serves a distinct purpose. A change in the number of blood cell types would result in an illness or disease. A low count of WBCs can cause various illnesses, including blood cancer. A lower count of healthy red blood cells leads to Anemia [5], and a lower ratio of platelets leads to excessive bleeding. Conventional blood cell type detection methods take a long time and have low accuracy, which highlights the significance of accurate systems for the rapid and precise analysis of blood cells [6]. Blood cells in microscopic images of blood smears have been categorized using conventional machine learning (ML) techniques like support vector machine, decision tree, k-nearest neighbor, naive Bayes, and artificial neural network [6,7]. The general process for traditional ML approaches includes pre-processing of blood smear images, segmentation to divide the cells, feature extraction, feature selection to remove undesired data, and classification. Despite many promising results, feature extraction and selection significantly affect how well classical ML algorithms perform in classification. Choosing the optimal features and finding the appropriate feature extraction algorithm has become complex and time-consuming [7]. Several deep learning (DL) techniques of convolutional neural networks (CNNs) have recently been proposed to tackle this challenging topic. Recent developments in deep learning allow us to estimate the type of blood cells from microscopic images. In contrast to conventional ML approaches, DL-based approaches have the capability of autonomous feature extraction and selection. Prior research revealed that for classifying blood cells, CNN performed better than traditional ML techniques [8].

The aim of this research is to develop a model for the classification of various blood cells, using machine learning methods. For this purpose, firstly, we have used transfer learning to evaluate the performance of CNN pre-trained models: VGG16, VGG19, ResNet-50, ResNet-101, ResNet-152, InceptionV3, MobileNetV2, and DenseNet-201 for microscopic image dataset provided by PBC dataset and evaluated their performance. Then we have introduced a CNN model for the classification of ten major blood subtypes. Our work aims to develop a convolution neural network (CNN) based model with decent generalization ability for the classification of various types of blood cells.

The paper is organized as follows. Section 2 introduces the related work. In Section 3, the blood dataset is presented. Section 4 explains the transfer learning approach. Moving on to Section 5, the proposed methodology is detailed. Results and their analysis are reported in Section 6, followed by a discussion. The last section draws the conclusions.

## II. RELATED WORK

Elhassan et al. proposed a two-step deep learning model to categorize atypical lymphocytes and immature WBCs [9]. The problem of an unbalanced distribution of WBCs in blood samples was addressed using a new method known as the "GT-DCAE WBC augmentation model," which is a hybrid model based on geometric transformation (GT) and a deep convolutional autoencoder (DCAE). A hybrid multi-classification model known as the "two-stage DCAE-CNN atypical WBC classification model" was created to divide atypical WBCs into eight categories. The model's average accuracy, sensitivity, and precision were 97%, 97%, and 98%, respectively. Ahmad et al. presented an improved hybrid method for optimum deep feature extraction using DenseNet201 and Darknet53 [10]. The dominant characteristics were then chosen using an entropy-controlled marine predator algorithm. (ECMPA). A public dataset of 5000 images of five distinct subtypes of WBCs was used. The system obtained an overall average accuracy of 99.9% while reducing the size of the feature vector by more than 95%. Singh et al. [11] suggested white blood cell classification using CNN and multiple optimizers such as SGD, Adadelta, and Adam with a batch size of 32 and 10 epochs. The best outcomes were obtained by using the Adam optimizer. The Adam optimizer yielded performance parameter values of 97% accuracy, 99% recall, and F1 score of 98%. Darrin et al. presented a video analysis method to automatically classify imbalance red blood cells from videos to monitor the status of sickle cell anemia patients [12]. The videos consisted of 6-100 frames. The convolutional neural network (CNN) model and a recurrent CNN were combined, and an accuracy of 97% with an F1-score of 0.94 was achieved. To solve this problem of blood cell classification, Rabul and Salam suggested Otsu's thresholding with Gray Level Co-occurrence Matrix (GLCM) features [13]. The R, G, and B channels from the original RGB image were separated from the Kaggle BCCD public data set to conduct image subtraction between the blue-red and blue-green channels. Then, WBCs are extracted from B-G channels using Otsu's thresholding and morphological filtering, and features are decorrelated using an ANOVA test and a zero-phase component analysis (ZCA) whitening procedure. KNN classification produced an accuracy of 94.25%. Milkisa et al. proposed a training strategy for neural networks to highlight the malaria-infected red blood cell pixels using the NH malaria data set [14]. Masked images were used to highlight the diseased area by dividing the image into R, G, and B channels, and then the intensity of the red channel was increased. The proposed approach achieved an accuracy of 97.2%.

A method for classifying Multiple Myeloma (MM) and Acute Lymphoblastic Leukemia (ALL) using the SN-AM dataset was suggested by Deepika et al. [15]. With a minimal number of parameters and computation time, the model was trained using an optimized Dense Convolution Neural Network framework capable of identifying the type of cancer present in cells with a precision of 97.2%.

Ansari *et al.* attempted to create a deep learning model with a customized architecture for identifying acute leukemia using images of lymphocytes and monocytes. A unique dataset with images of acute lymphoblastic and acute myeloid leukemia (AML) is used, and a new dataset has been developed. A Generative Adversarial Network (GAN) increased the dataset's size [16]. Six convolution layers, four dense layers, and a SoftMax activation function were part of the proposed CNN model based on the Tversky loss function for categorizing acute leukemia images. The proposed model had a 99.5% accuracy rate for identifying the various kinds of acute leukemia.

To identify various blood cells in microscopic blood images, Alkafrawi and Ismail suggested an AlexNet-based classification model [17]. Using convolution neural networks, the tests were carried out on a dataset of 17,000 blood smear samples obtained from the Hospital Clinic of Barcelona. Five convolutional layers, three maximum pooling layers and three fully connected layers make up the AlexNet model. The AlexNet-based model had a minimal Quadratic Loss of 0.0049 and a high accuracy of 95.08%.

Lee et al. created a new CNN-based blood cell detection and counting architecture [18]. VGG-16 was used to generate feature maps, which were then enhanced by feature fusion and a convolutional block attention mechanism (CBAM). The experiments on detecting RBCs, WBCs, and platelets were conducted using the BCCD dataset with two levels of certainty: 0.9 and 0.8, respectively. CBAM, which enlarges input images 1.5 times and uses images in RGB and grayscale color spaces, achieved the best recalls for RBC detection: 82.3% and 86.7% under two confidence levels. Meanwhile, it achieved a precision of 74.7% and 70.1%. Region of interest, which performs image preprocessing and uses RBG and grayscale images, outperforms other models for WBC detection. Precision and recall are 76.1% and 95%, respectively.

Kareem et al. classified blood cells using two distinct scenarios, the first using CNN directly and the second using SVM [19]. A data collection containing 10295 cell images was used. CNN obtained an accuracy of 98.4%, while SVM achieved an accuracy of 90.6%.

Miserlis et al. proposed an AI-based diagnosis for Peripheral Arterial Disease (PAD) and created 11 different ANN models [20]. DenseNet201, ResNet50v2, EfficientNetB0, and EfficientNetB7 achieve 97.22% precision. Training and testing were carried out between 2 and 8 seconds. The EfficientNetB0 and Resnet50v2 networks exhibited the best accuracy and speed execution.

Arif et al. proposed a framework for automatic leukemia detection based on deep learning [21]. The framework comprises several layers, including convolutional layers, batch normalization, leaky ReLU, and max pooling layers, and a CNN model called AlexNet was used to identify leukemia. The proposed framework accurately classified the images as either normal or leukemia-affected. They achieved 98.05% accuracy, 97.59% specificity, 100% recall, and 99.06% F1 score.

Meena Devi and Neel Ambary [22] tested different CNN architectures to detect and classify WBC. To train CNN, AlexNet, VGG16, GoogleNet, and ResNet50 were tested and analyzed at the convolution layer. The VGG16 CNN architecture trained with transfer learning outperformed with an accuracy of 97.16% in detecting monocytes 98.40%, basophils 98.48%, lymphocytes 99.52%, eosinophils 96.5%, and neutrophils 95.05%. Relevant data samples must be generated to address the challenge of imbalanced data, incomplete data samples, and missing labels. Pandya et al. [23] described generating data samples using a deep convolutional generative adversarial network (DCGAN). The testing shows that the model generates WBC blood cell images with 99.44% accuracy.

Alnawayseh et al. suggested differential counting of white blood cells (WBCs) to assess the immune system state of a patient [24]. Raw pixels from the data collection were used as input, extraneous pixels were removed, A CNN model with three layers: a convolutional layer, a downsized pooling layer, and fully connected hidden layers was trained. Weights are automatically given to data sets based on loss and accuracy.

You Only Look Once version 5 (YOLOv5) was used by Luong et al. to suggest a method for classifying and counting white blood cells for the diagnosis of blood related diseases [25]. The blood cells were carefully labeled with 619 leukemia cells, 115 neutrophils, 80 lymphocytes, 23 eosinophils, and 73 monocytes and were accurately detected and classified with 93% accuracy by the YOLOv5 algorithm.

To automatically extract RBC features, Z. Liao and Y. Zhang suggested an ultrasonic RF signal convolutional neural network [26]. RBCA-VGG10, a network model with removed layers and modified VGG16 structure, beat LeNet, AlexNet, GoogleNet, and ResNet models in terms of accuracy by 10.15%, 9.30%, 4.40%, and 8.34%, respectively. In [27], the authors described a blood cell image classification algorithm based on Efficient Net that used EfficientNet-B7 as the classification model and Contrast Limited Adaptive Histogram Equalization (CLAHE) to enhance image quality during data preprocessing. The algorithm had a 99.6% accuracy rate. LeukoX is a technique developed to identify and categories WBCs based on physical characteristics [28]. The Least Entropy Combiner (LEC) network combined the individual classifier results. Individual class results were fed to the proposed LEC network for learning. With an accuracy of 96.67%, the modal outperformed individual networks, with kappa and Matthew's correlation coefficient (MCC) values of 0.9334 and 0.9550, respectively.

According to the literature, the classification of blood cells is widely discussed. A large amount of research work has been done focusing on image classification and segmentation. Although only a few researchers have chosen to use manually crafted features for classification purposes. The earlier methods of classifying blood cells involved several steps such as preprocessing, selecting relevant features, and extracting meaningful information. In recent times, there is a growing trend to use convolutional neural networks (CNNs) to enhance the performance of classifying various types of blood cells.

## III. BLOOD CELL DATASET

### A. PBC dataset normal DIB

The PBC dataset [29] includes 17,092 images of different normal cells, shown in Table 1, that were collected using the CellaVision DM96 analyzer in the Core Laboratory of the Hospital Clinic of Barcelona. The dataset is divided into the following eight categories: Neutrophils, eosinophils, basophils, lymphocytes, monocytes, immature granulocytes (promyelocytes, myelocytes, and metamyelocytes), erythroblasts, and platelets or thrombocytes shown in Figure 1. Experienced clinical pathologists labeled the images and have a size of 360 x 363 pixels in the jpg format. The people whose images were taken during blood collection were free of infection, hematologic, or oncologic disease and were not taking any pharmaceutical treatment. The different types of typical peripheral blood cells may be recognized using this high-quality labeled dataset, which can be used to train and evaluate deep learning and machine learning models.

Table 1: Each group's cell types and numbers.

| Type of Cell | Images total by type |
| --- | --- |
| Neutrophils | 3329 |
| Eosinophils | 3117 |
| Basophils | 1218 |
| Lymphocytes | 1214 |
| Monocytes | 1420 |
| Immature granulocytes (Metamyelocytes, Myelocytes and Promyelocytes) | 2895 |
| Erythroblasts | 1551 |
| Platelets (Thrombocytes) | 2348 |
| Total | 17,092 |

(a) (b) (c) (d)

(e) (f) (g) (h)

(i) (j)

Figure 1. The dataset includes images of normal peripheral blood cells organized into eight groups. These groups cover cells commonly seen in infections and regenerative anemias.

### B. Components

The proposed work in Python utilizes various tools such as TensorFlow, Pandas, NumPy, Matplotlib, and Scikit-learn. The entire project is conducted using Google Collaboratory.

## IV. TRANSFER LEARNIGN

Transfer learning is a valuable machine learning technique that enables us to utilize an existing model, initially created for one task, to solve a different task. This method brings several advantages by saving time and resources that would otherwise be needed to construct neural network models from the ground up. It is especially useful in computer vision and natural language processing fields, where it enhances the accuracy and performance of the resulting model.

In this study, transfer learning has been employed by utilizing well-known CNN models, including VGG16, VGG19, ResNet50, ResNet101, ResNet-152, InceptionV3, MobileNetV2 and DenseNet201. This approach enables leveraging the pre-trained weights and architectures of these models, leading to substantial reductions in training time while enhancing the accuracy of the model.

VGG16 [30] is a deep convolutional neural network known for its significant contributions to image classification. Proposed by researchers Karen Simonyan and Andrew Zisserman, VGG16 consists of 16 weight layers, including 13 convolutional layers, five max-pooling layers, and three fully connected layers. The network utilizes small receptive fields of 3x3 for uniform feature extraction. ReLU activation functions introduce non-linearity, while max pooling layers reduce spatial dimensions. The final layers include three fully connected layers, with SoftMax activation for class probabilities.

VGG19 [31] is a convolutional neural network with 19 layers. It features a simple, uniform design using 3x3 convolutional filters and 2x2 max pooling layers. VGG19's stacked convolutional layers learn complex patterns and hierarchical representations. The final fully connected layers generate class probabilities. Despite its simplicity, VGG19 performs remarkably in image classification and is a popular baseline model in computer vision research.

ResNet50 [32] is a robust deep convolutional neural network architecture proposed by Kaiming He, Xiangyu Zhang, Shaoqing Ren, and Jian Sun at Microsoft Research. It revolutionizes image classification and feature extraction tasks by using residual connections. With 50 weight layers, including convolutional, bottleneck, and fully connected layers, ResNet50 effectively captures image features. The architecture employs 3x3 filters, bottleneck layers, and residual connections to enable gradient flow, address the vanishing gradient problem, and train deeper models. The final fully connected layers generate class probabilities.

ResNet-101 [33] is a deep neural network with 101 layers, including convolutional, pooling, and fully connected layers. It addresses the challenges of training deep networks using residual blocks with skip connections. These blocks facilitate gradient flow and mitigate the vanishing gradient problem. ResNet-101 employs bottleneck layers to reduce computational complexity while preserving representation capacity. It consists of multiple stages with varying numbers of residual blocks, enabling the extraction of hierarchical features.

ResNet-152 [34] is a deep convolutional neural network with 152 layers. It employs skip connections in residual blocks to tackle training challenges in deep networks. The architecture includes bottleneck layers for efficiency and multiple stages with varying numbers of residual blocks to extract hierarchical features. The final fully connected layers generate class probabilities for image recognition. With its depth and feature-capturing capabilities, ResNet-152 achieves superior performance, enabling accurate and robust deep neural networks.

The InceptionV3 [35] model is a deep convolutional neural network architecture developed by researchers at Google Research. Renowned for its innovative use of inception modules, InceptionV3 employs parallel convolutional layers with different filter sizes (1x1, 3x3, 5x5) and pooling operations to capture information at multiple scales and resolutions. The model incorporates batch normalization layers for efficient training and gradient flow. With a composition of multiple inception modules, fully connected layers, and a SoftMax activation function, InceptionV3 leverages pretraining on datasets like ImageNet to learn hierarchical representations and exhibit strong generalization capabilities.

MobileNetV2 [36] is an efficient convolutional neural network for mobile and embedded vision applications. It achieves a balance between model size and accuracy using depth wise separable convolutions and linear bottlenecks. By employing inverted residual blocks with residual connections, batch normalization layers, and ReLU6 activations, MobileNetV2 captures and propagates information effectively. Pretrained on ImageNet, it learns expressive features and demonstrates state-of-the-art performance on mobile devices. Its lightweight architecture makes it suitable for real-time applications in resource-constrained environments.

DenseNet201[37] is a deep convolutional neural network with 201 layers. It employs dense connectivity, connecting each layer to every other layer for efficient information flow and feature reuse. It includes convolutional, pooling, and fully connected layers and bottleneck layers for reduced computational complexity. DenseNet201 achieves high accuracy and improved gradient flow in computer vision tasks, making it valuable for image recognition and feature extraction applications.

## V. PROPOSED CNN MODEL

The proposed CNN model consists of 22 layers, divided into eight convolution blocks responsible for feature extraction, followed by fully connected layers (dense layers) for classification purposes. This architecture is specifically designed to process RGB images and aims to train input images with dimensions of (360 x 363), enabling the classification of ten distinct classes.

Each convolution block consists of a convolution layer, a max pooling layer, and a dropout layer. In a deep CNN, the convolutional layers apply filters to the original image or other feature maps.

When an input undergoes the same filter, it generates a feature map by performing a convolution operation on the input and passing the output to the subsequent layer. Consequently, a convolution layer performs pixel-wise multiplication between a two-dimensional input array (image) and a two-dimensional weight array (kernel).

Since the operation performed is a dot product, it yields a single value for each multiplication. However, due to the repetitive application of the filter to the input array, the resulting output is a two-dimensional array called a feature map. The purpose of employing filters is to identify specific features within the input and systematically scan the entire input image using these filters. This enables the filters to detect those particular features throughout the image.

The technique of pooling layers reduces the sampling of feature maps by summarizing the presence of features in specific map regions. By incorporating pooling layers, the size of the feature maps is reduced. More specifically, after applying a non-linear function like ReLU to the feature maps generated by a convolutional layer, the pooling layer operates independently on each feature map. This process leads to the creation a new set of pooled feature maps.

The dropout technique is utilized to prevent overfitting in a model. In a neural network, excessive weights indicate a more complex network that may overfit the training data. Dropout is a simple yet effective regularization approach by randomly removing nodes from the network during training. When applying dropout, employing a larger network with abundant training data and considering incorporating weight constraints is recommended.

After evaluating the performance of eight pre-trained models on the PBC dataset normal DIB, we found that these models did not achieve satisfactory accuracy for all blood cell classifications. To address this, we propose our convolutional neural network (CNN) architecture, shown in Figure 2. Our architecture comprises eight convolutional layers, eight pooling layers, five fully connected hidden layers, and an output layer.

The eight convolutional blocks in our architecture comprise convolutional (CONV) layers. These layers apply 3x3 convolutions with a stride of 1 and 'same' padding. They are followed by the Rectified Linear Unit (ReLU) activation function, which introduces non-linearity. Subsequently, MAXPOOL layers are used for 2x2 max pooling with a stride of 1. We also incorporate dropout layers with a dropout rate of 0.25 to improve the model's generalization ability.

### A. Convolutional Layer.

As previously mentioned, our architecture consists of eight convolutional layers, which are as follows:

1) First Layer: Kernel size: 3 x 3, number of filters: 32, Activation function: ReLU, Stride: 1, and input size: 100 x 100 (3 channels).
2) Second Layer: Kernel size: 3 x 3, number of filters: 64, Activation function: ReLU, Stride: 1
3) Third Layer: Kernel size: 3 x 3, number of filters: 64, Activation function: ReLU, Stride: 1.
4) Four Layer: Kernel size: 3 x 3, number of filters: 128, Activation function: ReLU, Stride: 1
5) Five Layer: Kernel size: 3 x 3, number of filters: 256, Activation function: ReLU, Stride: 1
6) Six Layer: Kernel size: 3 x 3, number of filters: 256, Activation function: ReLU, Stride: 1
7) Seven Layer: Kernel size: 3 x 3, number of filters: 256, Activation function: ReLU, Stride: 1
8) Eight Layer: Kernel size: 3 x 3, number of filters: 512, Activation function: ReLU, Stride: 1

### B. Pooling Layer.

Our architecture utilized max pooling with the same parameters for all eight pooling layers. The pooling layers were configured as follows: Pooling type: Maximum, Pooling Size: 2 x 2, Stride: 1, Dropout: 0.25.

### C. Fully Connected Layer.

In our CNN model, the final layers consist of fully connected layers. Our proposed methodology includes five fully connected hidden layers and one fully connected output layer. Here is the breakdown of the layers:

1) Fully Connected Hidden Layer: Total nodes: 128, Activation: ReLU.
2) Fully Connected Output Layer: Total nodes: number of classes, Activation: SoftMax.

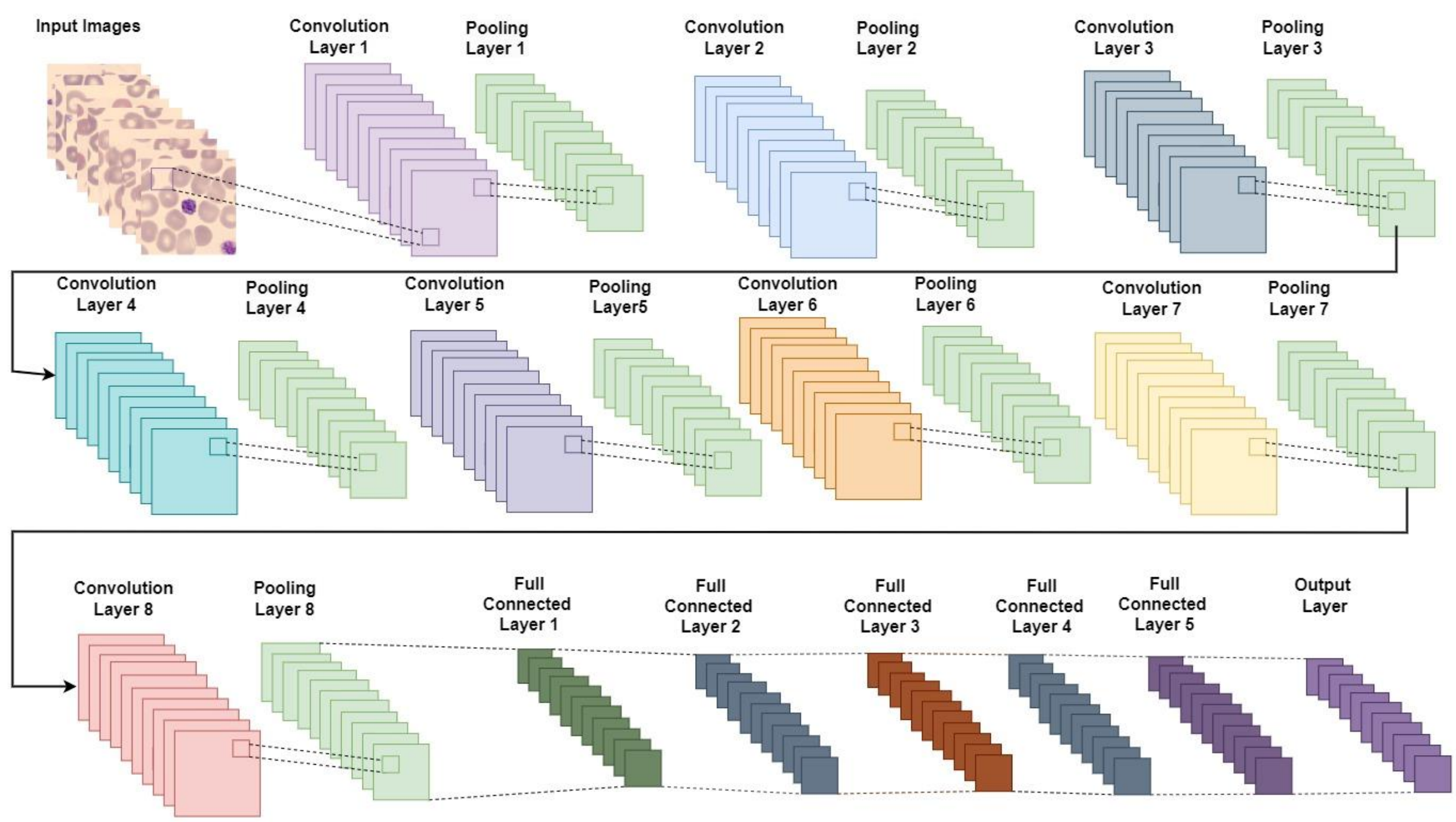

Figure 2 Architecture of our proposed CNN model

## VI. EXPERIMENTATION, RESULTS AND DISCUSSION

This section presents the experimental results, a comprehensive analysis, and a discussion of our findings.

### A. Evaluation Parameters

Machine learning models are evaluated based on specific parameters that measure their performance. This study employs four commonly used parameters, accuracy, recall, precision, and F-measure, to assess the model's fitness.

### B. Pre-Trained Model Compilation and Results

In the first phase of the evaluation, several pre-trained architectures, including VGG16, VGG19, ResNet50, ResNet101, ResNet-152, Inception V3, MobileNetV2 and DenseNet201 were assessed for the normal DIB of the PBC dataset. These architectures were trained using sparse categorical cross-entropy loss with Adam as the gradient-based optimizer. The pre-trained weights utilized were derived from the ImageNet dataset classification. The training focused on the last dense layers and spanned 150 epochs, with a learning rate 0.001.

The performance parameters achieved with the pre-trained architectures are summarized in Table 2 for the normal DIB of the PBC dataset.

With the VGG-16 architecture, 100 images out of 17,092 were misclassified. Specifically, there were misclassifications in the following categories: 15 images from eosinophils, 11 images from lymphocytes, 13 images from monocytes, 13 images from neutrophils, nine images from basophils, 15 images from immature granulocytes (metamyelocytes, myelocytes, and promyelocytes), 11 images from erythroblasts, and 13 images from platelets (thrombocytes). The overall accuracy achieved with VGG-16 is 92.8%. With the VGG-19 architecture, a total of 88 images were misclassified. The misclassifications include 14 images from eosinophils, 17 images from lymphocytes, eight images from monocytes, 12 images from neutrophils, eight images from basophils, ten images from immature granulocytes, eight from erythroblasts, and 11 from platelets. The overall accuracy achieved with VGG-19 is 91.8%.

For the ResNet-50 architecture, a total of 88 images were misclassified. These include 34 images from eosinophils, six images from lymphocytes, 13 images from monocytes, 14 images from neutrophils, 33 images from basophils, 15 images from immature granulocytes, 15 images from erythroblasts, and 16 images from platelets. The overall accuracy achieved with ResNet-50 is 94.72%. Using the ResNet-101 architecture, a total of 83 images were misclassified. These include 29 images from eosinophils, seven images from lymphocytes, 11 images from monocytes, 13 images from neutrophils, 30 images from basophils, nine images from immature granulocytes, 12 images from erythroblasts, and 11 images from platelets. The overall accuracy achieved with ResNet-101 is 93.7%. With the ResNet-152 architecture, a total of 85 images were misclassified. These include 15 images from eosinophils, four images from lymphocytes, seven images from monocytes, 18 images from neutrophils, 20 images from basophils, six images from immature granulocytes, eight images from erythroblasts, and ten images from platelets. The overall accuracy achieved with ResNet-152 is 91.375%.

Using the InceptionV3 architecture, a total of 126 images were misclassified. These include 17 images from eosinophils, seven images from lymphocytes, 20 images from monocytes, 14 images from neutrophils, 17 images from basophils, 16 images from immature granulocytes, 11 images from erythroblasts, and 31 images from platelets. The overall accuracy achieved with InceptionV3 is 93.125%. With the MobileNetV2 architecture, a total of 169 images were misclassified. These include 15 images from eosinophils, nine images from lymphocytes, 19 images from monocytes, 20 images from neutrophils, 18 images from basophils, 30 images from immature granulocytes, 16 images from erythroblasts, and 34 images from platelets. The overall accuracy achieved with MobileNetV2 is 92.01%.

For the DenseNet201 architecture, a total of 80 images were misclassified. These include 15 images from eosinophils, four images from lymphocytes, 18 images from monocytes, seven images from neutrophils, nine images from basophils, 11 images from immature granulocytes, seven images from erythroblasts, and 16 images from platelets. The overall accuracy achieved with DenseNet201 is 94.262%.

Table 2. Test results of pre-trained models on PBC dataset normal DIB

| Model | Type | Truth | Classified | Accuracy (%) | Precision (%) | Recall (%) | F-measure (%) |
|---|---|---|---|---|---|---|---|
| **VGG16** | Eosinophils | 3102 | 3117 | 0.930 | 0.93 | 0.82 | 0.87 |
| | Lymphocytes | 1203 | 1214 | 0.961 | 0.96 | 0.97 | 0.96 |
| | Monocytes | 1407 | 1420 | 0.950 | 0.95 | 0.98 | 0.96 |
| | Neutrophils | 3316 | 3329 | 0.94 | 0.93 | 0.97 | 0.95 |
| | Basophils | 1209 | 1218 | 0.90 | 0.91 | 0.81 | 0.85 |
| | Immature granulocytes (Metamyelocytes, Myelocytes and Promyelocytes) | 2880 | 2895 | 0.92 | 0.92 | 0.82 | 0.86 |
| | Erythroblasts | 1540 | 1551 | 0.91 | 0.92 | 0.81 | 0.85 |
| | Platelets (Thrombocytes) | 2335 | 2348 | 0.92 | 0.931 | 0.825 | 0.87 |
| **VGG19** | Eosinophils | 3103 | 3117 | 0.94 | 0.94 | 0.99 | 0.96 |
| | Lymphocytes | 1197 | 1214 | 0.92 | 0.92 | 0.80 | 0.86 |
| | Monocytes | 1412 | 1420 | 0.97 | 0.97 | 0.99 | 0.98 |
| | Neutrophils | 3219 | 3329 | 0.86 | 0.86 | 0.90 | 0.88 |
| | Basophils | 1210 | 1218 | 0.90 | 0.90 | 0.801 | 0.84 |
| | Immature granulocytes (Metamyelocytes, Myelocytes and Promyelocytes | 2885 | 2895 | 0.90 | 0.91 | 0.80 | 0.83 |
| | Erythroblasts | 1543 | 1551 | 0.90 | 0.90 | 0.802 | 0.84 |
| | Platelets | 2337 | 2348 | 0.961 | 0.96 | 0.97 | 0.96 |
| | (Thrombocytes) | | | | | | |
| **RsesNet50** | Eosinophils | 3083 | 3117 | 0.942 | 0.94 | 0.92 | 0.93 |
| | Lymphocytes | 1208 | 1214 | 0.982 | 0.98 | 0.99 | 0.98 |
| | Monocytes | 1407 | 1420 | 0.972 | 0.97 | 0.99 | 0.98 |
| | Neutrophils | 3315 | 3329 | 0.932 | 0.93 | 0.93 | 0.93 |
| | Basophils | 1185 | 1218 | 0.95 | 0.94 | 0.81 | 0.85 |
| | Immature granulocytes (Metamyelocytes, Myelocytes and Promyelocytes) | 2880 | 2895 | 0.930 | 0.93 | 0.82 | 0.87 |
| | Erythroblasts | 1536 | 1551 | 0.930 | 0.93 | 0.82 | 0.87 |
| | Platelets (Thrombocytes) | 2332 | 2348 | 0.94 | 0.94 | 0.81 | 0.88 |
| **ResNet-101** | Eosinophils | 3088 | 3117 | 0.92 | 0.89 | 0.85 | 0.90 |
| | Lymphocytes | 1207 | 1214 | 0.971 | 0.97 | 0.97 | 0.97 |
| | Monocytes | 1409 | 1420 | 0.952 | 0.95 | 0.98 | 0.97 |
| | Neutrophils | 3316 | 3329 | 0.932 | 0.93 | 0.88 | 0.90 |
| | Basophils | 1188 | 1218 | 0.92 | 0.92 | 0.825 | 0.87 |
| | Immature granulocytes (Metamyelocytes, Myelocytes and Promyelocytes) | 2886 | 2895 | 0.90 | 0.91 | 0.81 | 0.85 |
| | Erythroblasts | 1539 | 1551 | 0.94 | 0.95 | 0.97 | 0.96 |
| | Platelets (Thrombocytes) | 2337 | 2348 | 0.961 | 0.96 | 0.97 | 0.96 |
| **ResNet-152** | Eosinophils | 3102 | 3117 | 0.930 | 0.93 | 0.82 | 0.87 |
| | Lymphocytes | 1210 | 1214 | 0.94 | 0.95 | 0.95 | 0.95 |
| | Monocytes | 1413 | 1420 | 0.87 | 0.89 | 0.81 | 0.85 |
| | Neutrophils | 3311 | 3329 | 0.93 | 0.93 | 0.80 | 0.86 |
| | Basophils | 1198 | 1218 | 0.90 | 0.92 | 0.825 | 0.87 |
| | Immature granulocytes (Metamyelocytes, Myelocytes and | 2889 | 2895 | 0.87 | 0.89 | 0.80 | 0.86 |
| | Promyelocytes) | | | | | | |
| | Erythroblasts | 1543 | 1551 | 0.97 | 0.97 | 0.99 | 0.98 |
| | Platelets (Thrombocytes) | 2338 | 2348 | 0.90 | 0.91 | 0.80 | 0.83 |
| **InceptionV3** | Eosinophils | 3100 | 3117 | 0.90 | 0.90 | 0.84 | 0.85 |
| | Lymphocytes | 1207 | 1214 | 0.97 | 0.96 | 0.97 | 0.97 |
| | Monocytes | 1400 | 1420 | 0.90 | 0.92 | 0.82 | 0.84 |
| | Neutrophils | 3315 | 3329 | 0.94 | 0.95 | 0.98 | 0.96 |
| | Basophils | 1201 | 1218 | 0.91 | 0.90 | 0.83 | 0.87 |
| | Immature granulocytes (Metamyelocytes, Myelocytes and Promyelocytes) | 2879 | 2895 | 0.97 | 0.93 | 0.95 | 0.95 |
| | Erythroblasts | 1540 | 1551 | 0.94 | 0.96 | 0.95 | 0.97 |
| | Platelets (Thrombocytes) | 2317 | 2348 | 0.92 | 0.921 | 0.83 | 0.85 |
| **MobileNet V2** | Eosinophils | 3102 | 3117 | 0.90 | 0.90 | 0.81 | 0.83 |
| | Lymphocytes | 1205 | 1214 | 0.95 | 0.94 | 0.94 | 0.96 |
| | Monocytes | 1401 | 1420 | 0.90 | 0.90 | 0.80 | 0.81 |
| | Neutrophils | 3309 | 3329 | 0.92 | 0.93 | 0.96 | 0.94 |
| | Basophils | 1200 | 1218 | 0.91 | 0.92 | 0.80 | 0.85 |
| | Immature granulocytes (Metamyelocytes, Myelocytes and Promyelocytes) | 2865 | 2895 | 0.95 | 0.90 | 0.92 | 0.93 |
| | Erythroblasts | 1535 | 1551 | 0.93 | 0.94 | 0.92 | 0.94 |
| | Platelets (Thrombocytes) | 2314 | 2348 | 0.90 | 0.90 | 0.81 | 0.83 |
| **DenseNet201** | Eosinophils | 3102 | 3117 | 0.930 | 0.93 | 0.82 | 0.87 |
| | Lymphocytes | 1210 | 1214 | 0.98 | 0.981 | 0.99 | 0.98 |
| | Monocytes | 1402 | 1420 | 0.93 | 0.93 | 0.80 | 0.86 |
| | Neutrophils | 3322 | 3329 | 0.96 | 0.96 | 0.99 | 0.98 |

| | | | | | | |
|---|---|---|---|---|---|---|
| Basophils | 1209 | 1218 | 0.90 | 0.91 | 0.81 | 0.85 |
| Immature granulocytes (Metamyelocytes, Myelocytes and Promyelocytes) | 2884 | 2895 | 0.961 | 0.96 | 0.97 | 0.96 |
| Erythroblasts | 1544 | 1551 | 0.95 | 0.97 | 0.96 | 0.98 |
| Platelets (Thrombocytes) | 2332 | 2348 | 0.930 | 0.93 | 0.82 | 0.87 |

## V. Proposed CNN Model Results

To improve the performance of our model, we took inspiration from the existing architectures mentioned earlier and developed our own CNN model. We carefully considered three key parameters for training: the loss function, optimizer, and evaluation metrics. We utilized the sparse categorical cross-entropy loss function for our CNN model in conjunction with the widely used Adam optimizer. The training process involved feeding the training dataset to our model and training it for 150 epochs, with the best weights saved based on the loss function. Subsequently, we evaluated our proposed convolutional neural network model using all the blood cell images from the PBC dataset's normal DIB category.

### A. Results on PBC dataset normal DIB.

The graph in Figure 3 displays the model's loss and accuracy achieved during each epoch for the PBC dataset's normal DIB category. The network's performance is evaluated using the cross-entropy loss function, commonly employed to assess the effectiveness of convolutional neural networks. The cross-entropy value increases when the predicted value differs from the actual value. Ideally, the cross-entropy value should be zero.

In our case, we observed that the cross-entropy value reaches its minimum of 0.026 after 140 epochs, indicating a proximity to zero. The maximum error, which combines training and validation, is 0.057. Additionally, the training and validation accuracy of our convolutional neural network for the PBC dataset's normal DIB category is depicted in Figure 3. The highest training accuracy recorded is 0.993 after 142 epochs, while the maximum validation accuracy achieved is 0.985.

Using our proposed CNN model, we saved the model weights corresponding to the minimum loss and utilized them to predict labels for the testing dataset. The results are presented as a confusion matrix, as shown in Table 3. For the PBC dataset's normal DIB category, misclassifications were observed for one image each in the eosinophils and basophils classes. This can be attributed to the similarity in shape and size between these two cell types, as explained earlier. However, all images of the other eight classes were correctly classified with 100% accuracy.

Table 4 displays each class's accuracy, precision, recall and F-measure rates. Precision rates of 99%, 100%, 100%, 100%, 99.3%, 100%, 100%, and 100% were achieved for eosinophils, lymphocytes, monocytes, neutrophils, basophils, immature granulocytes (metamyelocytes, myelocytes, and promyelocytes), erythroblasts, and platelets (thrombocytes), respectively. The F-measure rates were 99.3% for eosinophils, 100% for neutrophils, 100% for lymphocytes, 100% for monocytes, 98% for basophils, 100% for immature granulocytes, 100% for erythroblasts, and 100% for platelets. Recall rates of 99.4%, 100%, 100%, 100%, 98.5%, 100%, 100%, and 100% were achieved for eosinophils, lymphocytes, monocytes, neutrophils, immature granulocytes, erythroblasts, and platelets, respectively. The misclassification of eosinophils and basophils can be attributed to their high similarity in size and shape. All images of lymphocytes, monocytes, neutrophils, immature granulocytes, erythroblasts, and platelets were correctly classified with 100% accuracy. The average accuracy achieved for all classes combined was 99.91%.

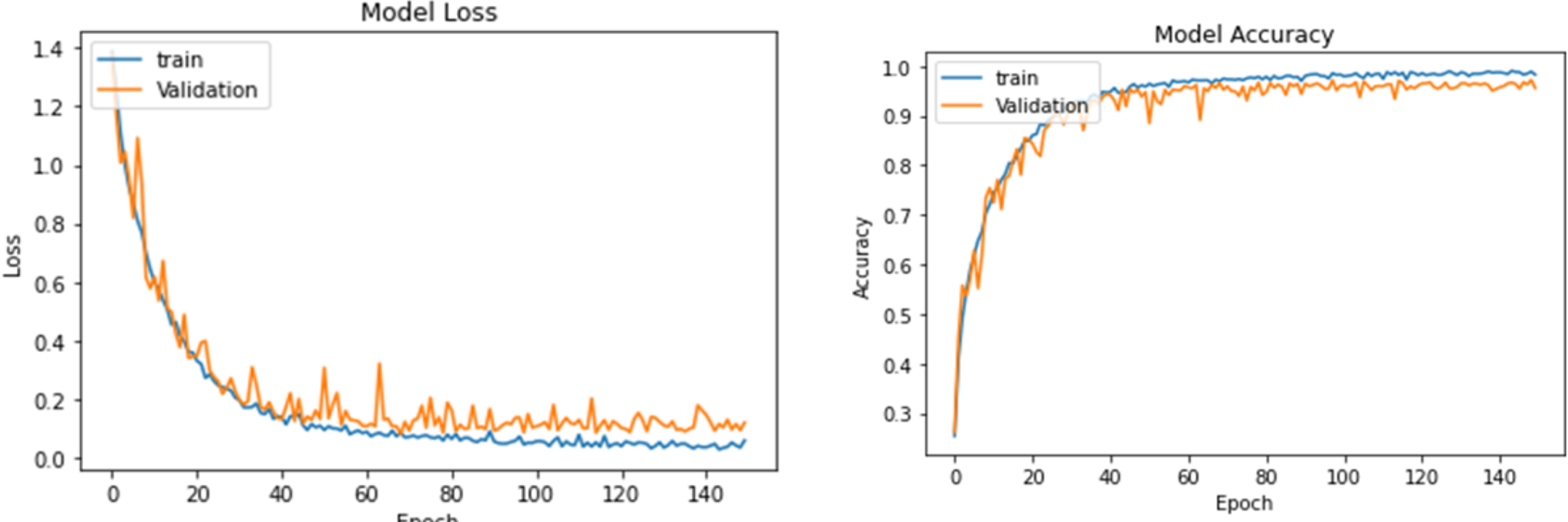

Figure 3. Graphs representing model loss and accuracy of proposed CNN model on PBC dataset

Table 3. Confusion matrix of P PBC dataset normal DIB

| Class | Eosinophil | Lymphocyte | Monocyte | Neutrophil | Basophils | Immature Granulocytes (Metamyelocytes, Myelocytes and Promyelocytes) | Erythroblasts | Platelets (Thrombocytes) |
|---|---|---|---|---|---|---|---|---|
| Eosinophil | 3116 | 0 | 0 | 0 | 0 | 0 | 1 | 0 |
| Lymphocyte | 0 | 1214 | 0 | 0 | 0 | 0 | 0 | 0 |
| Monocyte | 0 | 0 | 1420 | 0 | 0 | 0 | 0 | 0 |
| Neutrophil | 0 | 0 | 0 | 3329 | 0 | 0 | 0 | 0 |
| Basophils | 1 | 0 | 0 | 0 | 1217 | 0 | 0 | 0 |
| Immature Granulocytes (Metamyelocytes, Myelocytes and Promyelocytes) | 0 | 0 | 0 | 0 | 0 | 2895 | 0 | 0 |
| Erythroblasts | 0 | 0 | 0 | 0 | 0 | 0 | 1551 | 0 |
| Platelets (Thrombocytes) | 0 | 0 | 0 | 0 | 0 | 0 | 0 | 2348 |

Table 4. Test results of proposed CNN architecture

| Dataset | Type | Truth | Classified | Accuracy(%) | Precision (%) | Recall (%) | F measure (%) |
|---|---|---|---|---|---|---|---|
| | Eosinophils | 3116 | 3117 | 0.995 | 0.99 | 0.994 | 0.993 |
| | Lymphocytes | 1214 | 1214 | 100 | 100 | 100 | 100 |
| | Monocytes | 1420 | 1420 | 100 | 100 | 100 | 100 |
| | Neutrophils | 3329 | 3329 | 100 | 100 | 100 | 100 |
| | Basophils | 1217 | 1218 | 0.998 | 0.993 | 0.985 | 0.98 |

| | | | | | | |
|---|---|---|---|---|---|---|
| Immature granulocytes (Metamyelocytes, Myelocytes and Promyelocytes) | 2895 | 2895 | 100 | 100 | 100 | 100 |
| Erythroblasts | 1551 | 1551 | 100 | 100 | 100 | 100 |
| Platelets (Thrombocytes) | 2348 | 2348 | 100 | 100 | 100 | 100 |

Table 5. Comparison of proposed CNN architecture

| Author | Datasets | No. of Images | Model | Accuracy (%) | F1 score (%) | Recall (%) | Precision (%) |
|---|---|---|---|---|---|---|---|
| Prasenjit et. al[38] | PBC DIB | 17,092 | Watershed algorithm. | - | 97.95 | 98.1 | 97.81 |
| Prasenjit et. al[38] | ALL IDB1 | 108 | Watershed algorithm. | - | 96.15 | 96.1 | 96.2 |
| Prasenjit et. al[38] | Kaggle | 100 | Watershed algorithm. | - | 98.7 | 98.6 | 98.8 |
| Hilal Atici and Hasan Erdinc[39] | PBC DIB | 17,092 | ResNet101 | 0.9931 | 96.87 | 97.37 | 97.25 |
| Lubnaa Abdur Rahman and Poolan Marikannan Booma[40] | ABIDE | 2939 | MobileNet | 90 | 90 | 91 | 88 |
| Erdal Başaran [41] | Public dataset | 12435 | mRMR and LIME method with SVM(Feature size 400+5) | 95.15 | 93.13 | - | 95.13 |
| Tusneem et. al. [9] | Acquired from Munich University Hospital. | 18365 | | 97 | - | - | 97.42 |
| R. Ahmad et. al.[10] | Public dataset | 5000 | Dense Net and Darknet | 99.6 | - | - | - |
| Milkisa et. al. [14] | NIH Malaria dataset [14] | 19000 | ResNet and MobileNet | 99.53 | 97.00 | 94.90 | - |
| Proposed Methodology | PBC DIB | 17092 | CNN | 99.91 | 99.6 | 99.6 | 99.77 |

Table 5 shows the outcomes of our suggested method compared to other relevant works from the literature. Prasenjit et al [38] achieved a precision of 97.81% on the PBC DIB dataset, 96.2% on ALL DIB, and with the Kaggle dataset, the precision was 98.9%. Hilal Atici and Hasan Erdinc [39] used the PCB DIB dataset using ResNet101; they achieved an accuracy of 99.31%. Lubnaa Abdur Rahman and Poolan Marikannan Booma [40] used MobileNet on the ABIDE dataset to reach an accuracy of 90%. Erdal Başaran [41] used the WBC dataset with 12435 images and achieved an accuracy of 95.15%. R. Ahmad et al. [10] employed Dense Net and Darknet to achieve 99.6% accuracy on a public dataset of 5000 images. Tusneem et al. [9] obtained a 97% accuracy on a public dataset. Milkisa et al. [14] attained an accuracy of 99.53% utilizing the NIH Malaria dataset with ResNet and MobileNet. Compared to all previous work, our model achieves an accuracy of 99.91%, outperforming all previous work.

## CONCLUSION:

In this paper, we have first applied different pre-trained models, VGG16, VGG19, ResNet-50, ResNet-101, ResNet-152, InceptionV3, MobileNetV2, and DenseNet-201. Motivated by these architectures and a quest for better performance, a CNN-based model has been proposed to categorize the subtypes of blood cells (red blood cells, white blood cells, and platelets). The proposed architecture consists of eight convolutional layers, eight pooling layers, five fully connected hidden layers, and an output layer. The classification is performed using microscopic blood cell images obtained from the PBC dataset normal DIB. During testing, the proposed algorithm has shown optimal performance in terms of classification with 99.91% accuracy for the PBC dataset. The proposed model is effective as the results achieved are competitive in comparison with previous results reported in the literature on the same datasets. In the future, our proposed architecture can be applied to the classification of other cells and tissues of the body that can help the pathologists in effective diagnosis.

## REFERENCES:

[1] Tamang, T., Baral, S., & Paing, M. P. (2022). Classification of white blood cells: A comprehensive study using transfer learning based on convolutional neural networks. *Diagnostics (Basel, Switzerland)*, *12*(12).

*[2]* N. Lu *et al.*, "Label-free microfluidic cell sorting and detection for rapid blood analysis," *Lab Chip*, vol. 23, no. 5, pp. 1226–1257, 2023.

[3] F. Rustam *et al.*, "White blood cell classification using texture and RGB features of oversampled microscopic images," *Healthcare (Basel)*, vol. 10, no. 11, p. 2230, 2022.

[4] E. Gavas and K. Olpadkar, "Deep CNNs for peripheral blood cell classification," *arXiv [cs.CV]*, 2021.

[5] *M. Darrin, A. Samudre, M. Sahun, S. Atwell, C. Badens, A. Charrier , E. Helfer , A. Viallat, V. C. Addad & S. G. Roisin. "Classification of red cell dynamics with convolutional and recurrent neural networks: a sickle cell disease case study," Scientific Reports, 2023, 13 (1), pp.745.*

[6] S. Çelebi, 1 Histology and Embryology Department Doctoral Program, Yıldırım Beyazıt University, M. Burkay Çöteli, and 2 Graduate School of Informatics, Middle East Technical University, "Red and white blood cell classification using Artificial Neural Networks," *AIMS Bioeng.*, vol. 5, no. 3, pp. 179–191, 2018.

[7] T.-R. Tseng and H.-M. Huang, "Classification of peripheral blood neutrophils using deep learning," *Cytometry A*, 2022.

[8] Sahlol, P. Kollmannsberger, and A. A. Ewees, "Efficient classification of white Blood Cell Leukemia with improved swarm optimization of deep features," *Sci. Rep.*, vol. 10, no. 1, p. 2536, 2020.

[9] T. A. Elhassan *et al.*, "Classification of atypical white blood cells in acute myeloid leukemia using a two-stage hybrid model based on deep convolutional autoencoder and deep convolutional neural network," *Diagnostics (Basel)*, vol. 13, no. 2, p. 196, 2023.

[10] R. Ahmad, M. Awais, N. Kausar, and T. Akram, "White blood cells classification using entropy-controlled deep features optimization," *Diagnostics (Basel)*, vol. 13, no. 3, p. 352, 2023.

[11] R. Singh, A. Sharma, N. Sharma and R. Gupta, "Impact of Adam, Adadelta, SGD on CNN for White Blood Cell Classification," *2023 5th International Conference on Smart Systems and Inventive Technology (ICSSIT)*, Tirunelveli, India, 2023, pp. 1702-1709.

[12] Darrin, M., Samudre, A., Sahun, M. *et al.* Classification of red cell dynamics with convolutional and recurrent neural networks: a sickle cell disease case study. *Sci Rep* **13**, 745 (2023).

[13] R. Saikia and S. S. Devi, "White blood cell classification based on gray level co-occurrence matrix with zero phase component analysis approach," *Procedia Comput. Sci.*, vol. 218, pp. 1977–1984, 2023.

[14]M. Yebasse, K. J. Cheoi and J. Ko, "Malaria Disease Cell Classification With Highlighting Small Infected Regions," in *IEEE Access*, vol. 11, pp. 15945-15953, 2023

[15]D. Kumar *et al.*, "Automatic Detection of White Blood Cancer From Bone Marrow Microscopic Images Using Convolutional Neural Networks," in *IEEE Access*, vol. 8, pp. 142521-142531, 2020, doi: 10.1109/ACCESS.2020.3012292.

[16] S. Ansari , A. H. Navin, A. B. Sangar , J. V. Gharamaleki and S. Danishvar, "A Customized Efficient Deep Learning Model for the Diagnosis of Acute Leukemia Cells Based on Lymphocyte and Monocyte Image,*" Electeonics,*

[17] I. M. I. Alkafrawi and Z. A. Dakhell, "Blood cells classification using deep learning technique," in *2022 International Conference on Engineering & MIS (ICEMIS)*, 2022.

[18] S.-J. Lee, P.-Y. Chen, and J.-W. Lin, "Complete blood cell detection and counting based on deep neural networks," *Appl. Sci. (Basel)*, vol. 12, no. 16, p. 8140, 2022.

[19] H. A. Muhamad, S. W. Kareem and A. S. Mohammed, "A Comparative Evaluation of Deep Learning Methods in Automated Classification of White Blood Cell Images," *2022 8th International Engineering Conference on Sustainable Technology and Development (IEC)*, Erbil, Iraq, 2022

[20] D. Miserlis *et al.*, "Convolutional neural network analysis of tissue remodeling and myopathy in peripheral arterial disease," in *2022 13th International Conference on Information, Intelligence, Systems & Applications (IISA)*, 2022.

[21] R. Arif, S. Akbar, A. B. Farooq, S. Ale Hassan and S. Gull, "Automatic Detection of Leukemia through Convolutional Neural Network," *2022 International Conference on Frontiers of Information Technology (FIT)*, Islamabad, Pakistan, 2022, pp. 195-200.

[22] G. M. Devi and V. Neelambary, “Computer-aided diagnosis of white blood cell leukemia using VGG16 convolution neural network,” in *2022 4th International Conference on Inventive Research in Computing Applications (ICIRCA)*, 2022.
[23] D. Pandya, T. Patel, and D. K. Singh, “White blood cell image generation using deep convolutional generative adversarial network,” in *2022 International Conference on Augmented Intelligence and Sustainable Systems (ICAISS)*, 2022.
[24] S. E. A. Alnawayseh, W. T. Al-Sit, H. Alrababah, N. S. Yasin, M. Fatima and N. Mehmood, "Classification of White Blood Cells Empowered with Auto Encoder and CNN," *2022 International Conference on Cyber Resilience (ICCR)*, Dubai, United Arab Emirates, 2022, pp. 1-7.
[25] D. Trong Luong, D. Duy Anh, T. Xuan Thang, H. Thi Lan Huong, T. Thuy Hanh and D. Minh Khanh, "Distinguish normal white blood cells from leukemia cells by detection, classification, and counting blood cells using YOLOv5," *2022 7th National Scientific Conference on Applying New Technology in Green Buildings (ATiGB)*, Da Nang, Vietnam, 2022, pp. 156-160.
[26] Z. Liao and Y. Zhang, "Red Blood Cell Aggregation Classification Based on Ultrasonic Radiofrequency Echo Signals by An Improved Convolutional Neural Network," *2022 3rd International Conference on Computer Vision, Image and Deep Learning & International Conference on Computer Engineering and Applications (CVIDL & ICCEA)*, Changchun, China, 2022, pp. 195-198.
[27] Y. Guan and Z. Wang, “Blood cell image recognition algorithm based on EfficientNet,” in *2022 IEEE International Conference on Mechatronics and Automation (ICMA)*, Guilin, Guangxi, China, 2022, pp. 1640-1645,
[28] S. Ghosh, M. Majumder and A. Kudeshia, "LeukoX: Leukocyte Classification Using Least Entropy Combiner (LEC) for Ensemble Learning," in *IEEE Transactions on Circuits and Systems II: Express Briefs*, vol. 68, no. 8, pp. 2977-2981, Aug. 2021, doi: 10.1109/TCSII.2021.3064389.
[29] Acevedo, Andrea; Merino, Anna; Alférez, Santiago; Molina, Ángel; Boldú, Laura; Rodellar, José (2020), “A dataset for microscopic peripheral blood cell images for development of automatic recognition systems”, Mendeley Data, V1, doi: 10.17632/snkd93bnjr.1.
[30] K. Simonyan and A. Zisserman, “Very deep convolutional networks for large-scale image recognition,” 3rd Int. Conf. Learn. Represent. ICLR 2015 - Conf. Track Proc., pp. 1–14, 2015.
[31] V. N. Murthy, S. Maji, and R. Manmatha, “Automatic image annotation using deep learning representations,” in Proceedings of the 5th ACM on International Conference on Multimedia Retrieval, 2015.
[32] K. He, X. Zhang, S. Ren, and J. Sun, “Deep Residual Learning for Image Recognition,” in 2016 IEEE Conference on Computer Vision and Pattern Recognition (CVPR), Las Vegas, NV, USA, Jun. 2016, pp. 770–778.
[33] Y. Li et al., “KinNet: Fine-to-coarse deep metric learning for kinship verification,” in Proceedings of the 2017 Workshop on Recognizing Families In the Wild, 2017.
[34] Y. Li et al., “KinNet: Fine-to-coarse deep metric learning for kinship verification,” in Proceedings of the 2017 Workshop on Recognizing Families In the Wild, 2017.
[35] C. Szegedy, V. Vanhoucke, S. Ioffe, J. Shlens, and Z. Wojna, “Rethinking the Inception Architecture for Computer Vision,” in 2016 IEEE Conference on Computer Vision and Pattern Recognition (CVPR), Las Vegas, NV, USA, Jun. 2016, pp. 2818–2826.
[36] M. Sandler, A. Howard, M. Zhu, A. Zhmoginov, and L.-C. Chen, “MobileNetV2: Inverted Residuals and Linear Bottlenecks,” in 2018 IEEE/CVF Conference on Computer Vision and Pattern Recognition, Salt Lake City, UT, Jun. 2018, pp. 4510–4520.
[37] S.-H. Wang and Y.-D. Zhang, “DenseNet-201-based deep neural network with composite learning factor and precomputation for multiple sclerosis classification,” ACM Trans. Multimed. Comput. Commun. Appl., vol. 16, no. 2s, pp. 1–19, 2020.
[38] P. Dhar, K. Suganya Devi, S. K. Satti and P. Srinivasan, “Efficient detection and partitioning of overlapped red blood cells using image processing approach**”** Innovations in Systems and Software Engineering 2022.
[39] H. Atıcı and H. E. Kocer , "Mask R-CNN Based Segmentation and Classification of Blood Smear Images", *Gazi Journal of Engineering Science*, vol. 9, no. 1, pp. 128-143, Apr. 2023.
[40] L. A. Rahman and P. Marikannan Booma, "The Early Detection of Autism Within Children Through Facial Recognition; A Deep Transfer Learning Approach," *2022 2nd International Conference on New Technologies of Information and Communication (NTIC)*, Mila, Algeria, 2022, pp. 1-11, doi: 10.1109/NTIC55069.2022.10100517.
[41] E. Başaran, “Classification of white blood cells with SVM by selecting SqueezeNet and LIME properties by mRMR method ” Signal, Image and Video Processing 2022 Vol. 16 Issue 7 Pages 1821-1829. DOI: 10.1007/s11760-022-02141-2.